# The Hunt Hypothesis and the Dividend Policy of the Firm. The Chaotic Motion of the Profits

**Safieddine Bouali**
University of Tunis
Institut Supérieur de Gestion
41, rue de la Liberté-2000 Le Bardo, Tunisia
E-mail :Safieddine.Bouali@isg.rnu.tn

*Abstract :*

Presenting the best financial structure of the firm, Hunt (1961) introduces the concept of *trade on equity* where the call for loans pushes the profit ratio of stockholders'capital. This principle is obviously an accounting view of the profit ratio but its relevance constitutes only an hypothesis in dynamics. We have carried out the first (to our knowledge) simulations of firm model to analyse its validity in dynamics.
On the other hand, in our model, the distribution of dividend begins when the mass of profit exceeds a particular threshold chosen by the managers. Before this level, the braking of the profits withdrawal outside the firm drives a better self-financing. On the contrary, beyond the threshold, shared profits are not reinvested and the size of the firm (and its capital) may be reduced when the stockholders expect the downward side of the life cycle of the activity.
In this paper, we discuss and simulate a system of three ordinary differential equations gathering the variables of profits, reinvestments and financial flow of borrowings.
We investigate the implications of a greater running into debt associated with a more cautious policy of dividend distribution. With a nonlinear relation of reinvestments, the two connected policies exhibit the ability of the debt to inject perturbation into the profit motion. Indeed, the results display various periodicities and chaos. Nonetheless, a suplementary debt may amplify the level of probable losses. The same implication is observed when the dividend policy is more and more cautious. The 3D model shows a wide range of a dynamic behavior that may be trapped into chaotic attractors with strange characteristics.

> *« Static and dynamic […] are not two chapters belonging to the same theoretical corpus but represent two corpuses which are utterly different.»*
>
> **J. Schumpeter**

## I-Introduction

Many managers recommend the introduction of some borrowing proportion in order to reach the optimal financial structure of the firm [Myers, 1977]. The forerunner of the *Trade on Equity* principle was Hunt (1961) who showed that the rate of profitability of the shareholders' capital is all the more higher than entreprise finances its activities with debts. The success of this financial strategy is achieved when the profit ratio on total capitalization is greater than the interest rate of debt refunds [Myers, 1974 ; Glais, 1992].
*Ceteris paribus*, this massive borrowing will improve the profitability rate of the shares making up the shareholders' capital. Thus, the managers can dope the value of the firm in the stockmarket by increasing the capital with debts.
However, in dynamics, Hunt's principle constitutes only an hypothesis because the uncertainty about the profit amounts provide to the debt flow a multiplying effect on the profitability rate of the shareholders' capital.
Having recourse to debts in this case, is a lever that acts both on the positive and negative side of the profitability of the entreprise. Indeed, in case of a sharp growth, the profitability rate is enhanced by comparison to the entreprise being financed only by its shareholders' capital.
Yet, during a recession, the losses are multiplied and the firm greatly runs into debts.
On the other hand, reinforcing the financial means of a firm also goes through setting up a cautious policy of dividend distribution. Hence, lesser distribution contributes to a better profitability [Kalay, 1980]. Triggering the dividend distribution beyond a certain threshold of profit thereby represents a resource which will consolidate the finance structure of the firm [see Lintner, 1956; Fama and French, 1998, Fluck, 1999].
If this minimum level of profit is not accomplished, the shareholders' contribution is called to increase the capital of their firm.
The focus of this paper is to measure the profit trend when an entreprise chooses a gradual increase of its debt in relation to its capital so as to improve its profitability rate. In our model, the firm will also implement a cautious dividend policy based not on the rate, but the mass of profit itself as it will be able to boost its shareholders' capital.

## II. The model of the firm

Written in three first-order differential equations, our simplest and idealized model explores the firm's dynamics with basic questions of corporate financial policies of the firm.

Based on the first principles and rules of the best finance governance, our *ab initio* model provides a consistent dynamic exploration and experimentation which can serve to check the validy of these rules when a nonlinear system is chosen.

The theoretical approach which governs this work links reinvestment and dividend policies. To improve the analysis of these two connected policies, we built a 3D system that goes to the heart of which dynamic behavior of the profit arises in a finite number of numerical simulations.

### II-1. Model equations

In our 3D model of autonomous O.D.Es in which *dt* represents the year, all the variables are endogenous.

The capital of the firm is obviously at the origin of profit creation. It is made up of reinvestments R and financed by debts F.

Therefore, the 1/v coeficient represents the rate of profit:

$$dP/dt = (R + F)/v \qquad [1]$$

Moreover, the reinvestments are composed of a fraction of profits according the proportion m and of the capitalization of these very reinvestments that are reassessed yearly at rate n.

$$dR/dt = m\,P + n\,(1 - P^2)\,R \qquad [2]$$

Nevertheless, the distribution policy of the dividend interferes in this relation since the profit reaches exactly value 1, the nonlinear item becomes null and the yearly reinvestments shall remain at a constant trend m. In this figure case, the remainder of profits as well as the capital added value are distributed as dividends.

Yet, so long as the profit doesn't reach the unity threshold, the firm selects a cautious behavior to sustain its self-financing. The profits will be re-injected into the financing circuit of the firm. In fact, not only is the dividend distribution blocked, but also the capitalization increases at a rate more proportional than the gap separating 1 and P.

At such a stage, when the entreprise intends to achieve a faster increase by self-financing, there is a call for shareholders to contribute to a capital increase.

On the contrary, beyond the profit threshold ( P > 1), the capitalization is reduced by a great movement of dividend distribution.

For higher profit amounts, not only do reinvestments shrink, but also a reduction in shareholders' capital can be made.

The shareholders will therefore avoid the fall in the exchange rate of the stockmarket because these high profits herald a near decline in this activity and a tough competition that shall occur following the emergence of new firms.

They stop auto-financing and withdraw their funds as their firm's technology has become obsolete.

In the model, only the shared profits in this second equation follow a nonlinear mathematical specification.

Eventually, according to Hunt's hypothesis, the firm chooses an increase of its capital by borrowing according to the debt rate *s* proportional to self-financing.

$$dF/dt = -rP + sR \qquad [3]$$

The net capital inflow is obtained by deducting the refunding of the borrowals according to interest rate *r* to the profit amount.

We carried out the first (to our knowledge) simulations of O.D.Es model of firm to analyse the validity of Hunt's hypothesis when the dividend policy does not entail great expenses.

To identify the outcome of our model, the numerical results are computed with the fifth-order Runge-Kutta integration method and accuracy (or the unit of time) equal to $10^{-6}$. When not indicated, the simulations are started with initial conditions (P, R, F) = (0.01, 0.01, 0.01).

## II-2. Computational results

For the first simulation, we choose the set of parameters $P_0$ (v, m, n, r, s) = (4, 0.04, 0.02, 0.1, 0.2). The behavior of the state variables is a periodic motion (Fig. 1) wherein the orbits are centered around the unstable equilibrium (P, R, F) = (2.23, 1.11, -1.11).

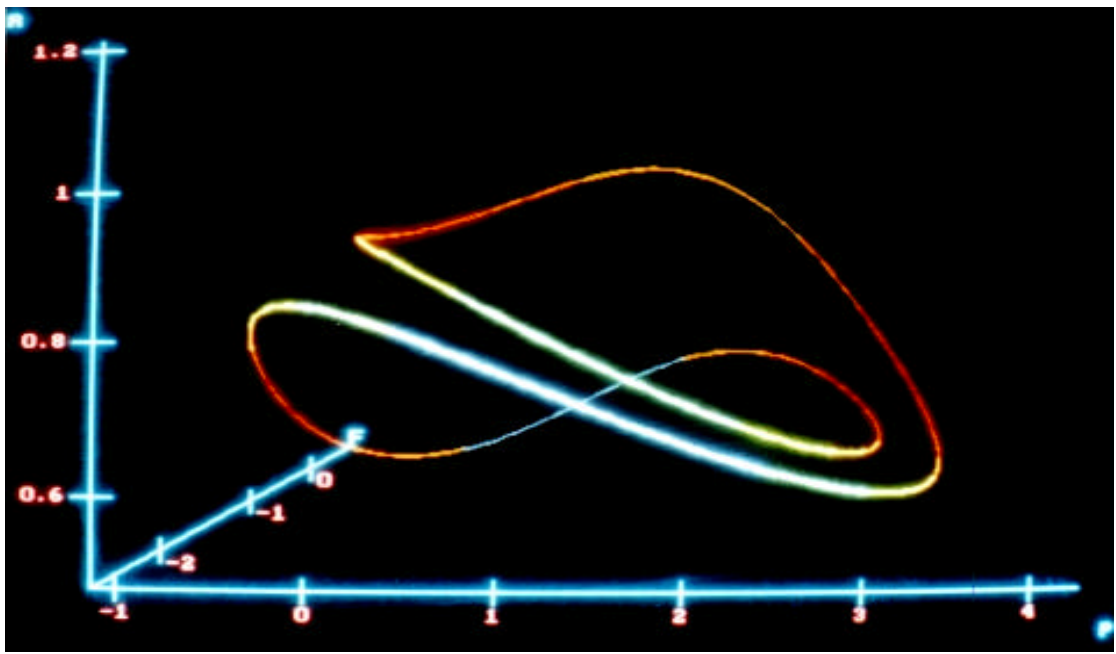

**Fig. 1.** *Period-4* solution for $P_0$.

If the firm chooses to gear up its debts (s = 0.3) to boost its activity, thereby improving the profit rate, it will undergo a chaotic dynamic (Fig. 2) of its variables, but this turbulence is contained in a field equivalent to the previous case (see Appendix).

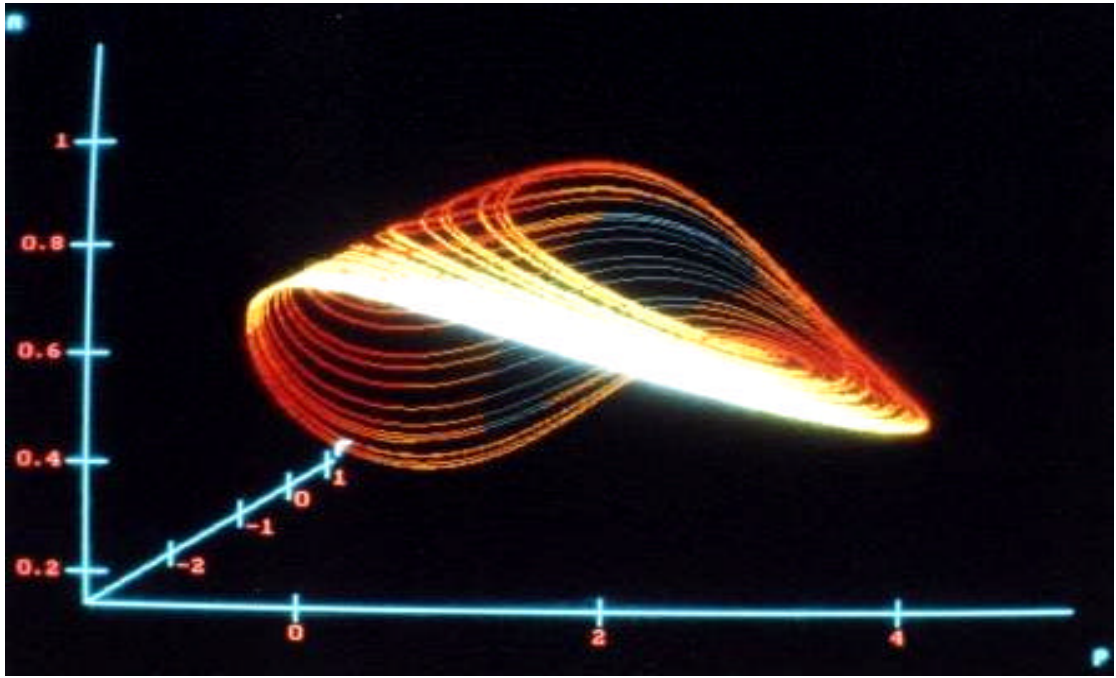

**Fig. 2. Chaotic attractor for P$_0$ when s = 0.3.**

Nonetheless, the intensification of this borrowal policy broadens the loss amplitude very acutely without reduction of the instability in the financial system of the firm. As a matter of fact, there is a hazardous chaotic expansion as shown in figure 3 with *sur-critiques* Hopf bifurcations.

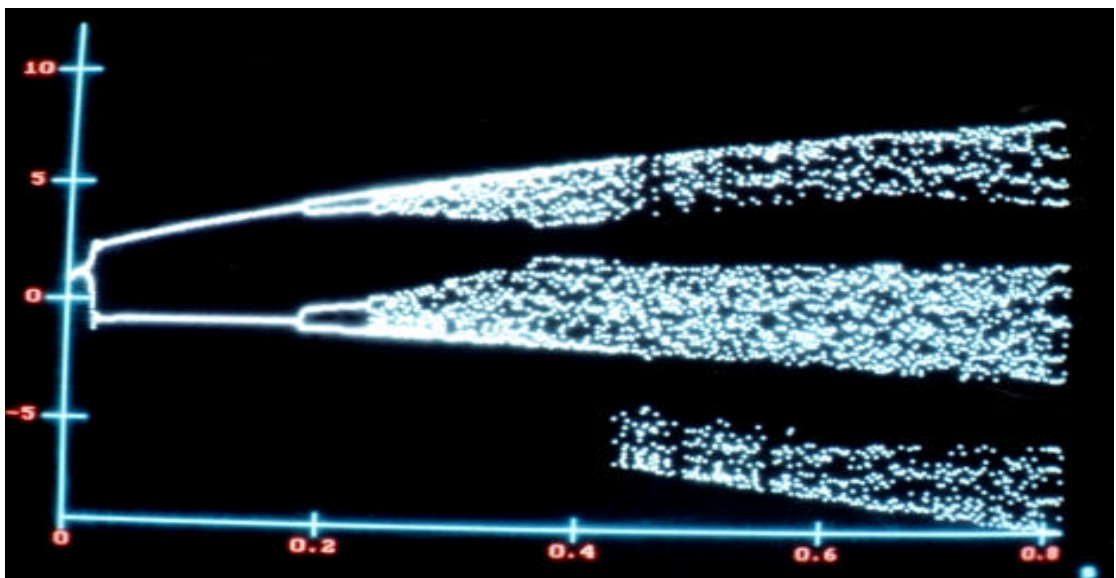

**Fig. 3. bifurcation diagram of profit ( s stepsize $10^{-5}$).** The borrowings implies the amplification of loss level

Beyond s = 0.4, the maximum level of losses reaches that of profits. For instance, at s = 0.6 an antisymmetric attractor (Fig. 4) occurs with an impressive decapitalization behavior ( negative values of R).

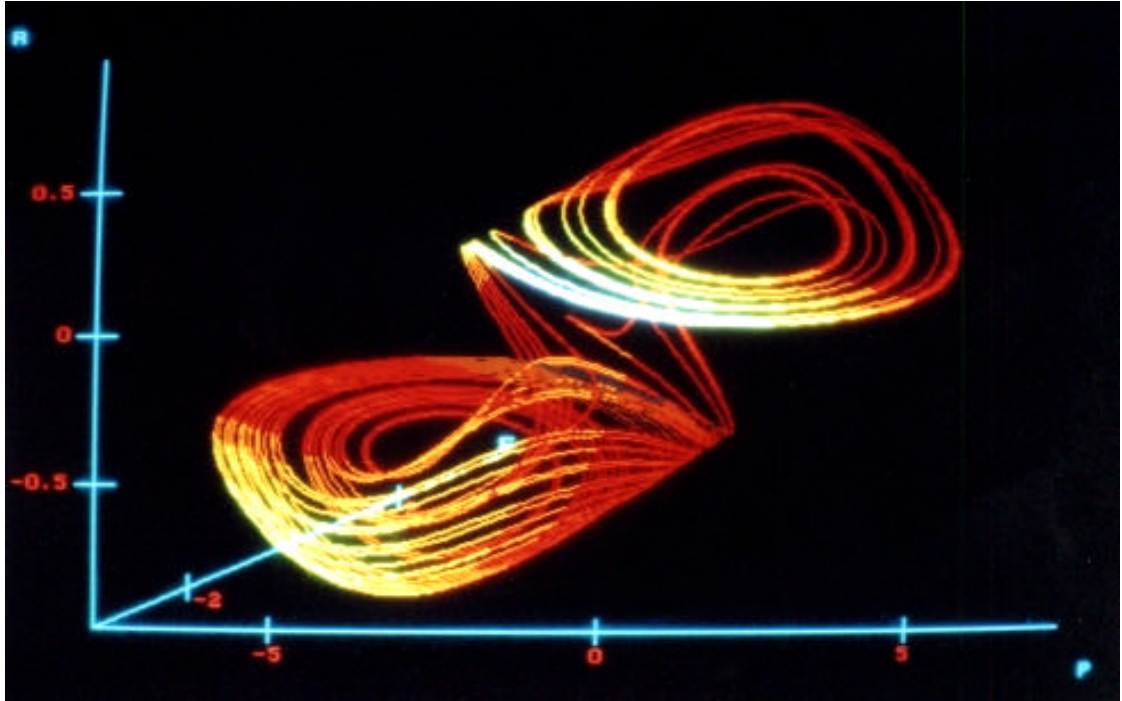

**Fig. 4. Chaotic attractor for $P_0$ when s = 0.6.**

These capital withdrawls accompanying the highest level of losses, represent a logical conjunction since the shareholders want to regain their funds at a rate higher than the losses. Worse is sR term which represents the borrowals and no longer finances the firm, but constitutes a capital flight.

To avoid the risks incurred by this borrowing policy, the entreprise can opt for a financial management which keeps the borrowals rate constant, for example s=0.3, but that curbs the dividend distribution. Thus, the minimum threshold that triggers the distribution will undergo a continous rise creating financial reserves for reinvestments.

With the new equation, where $dT/dt = 10^{-5}$ is the threshold trend:

$$dR/dt = m P + n ( T - P^2 )R \qquad [2.1]$$

and $T_0 = 1$, the dividend distribution is more and more cautious. However, this pattern does not reduce the risk of a negative profitability despite the retained earnings. In the neighborhood of T = 1.5, the possible worst performance moves sharply from -2 to - 6 (Fig. 5).

The fast-occuring profits are deviated outside the entreprise with accelerated motion in relation with the rise of the threshold.

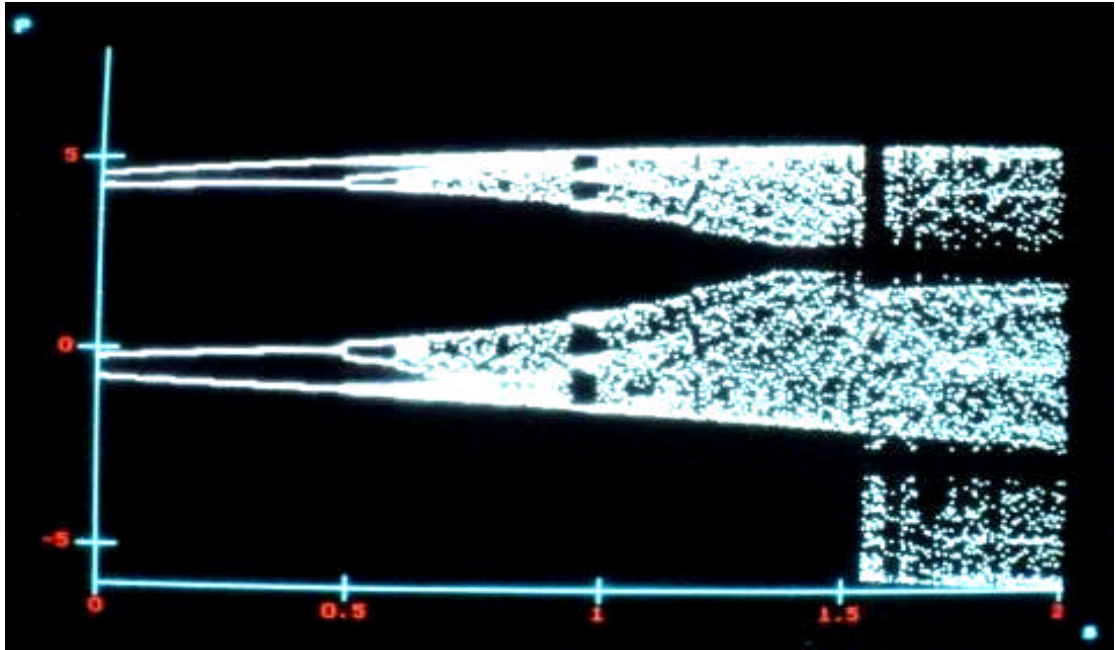

**Fig. 5. bifurcation diagram of profits ( T stepsize $10^{-5}$).** The cautious dividend policy implies the amplification of the loss level

### II-3. Theoretical foundations

The numerical simulations contain a wealth of information and a close inspection of the maps reveals the nonlinear characteristics of the dynamics. Indeed, the model produces a strong plasticity of chaotic attractors. For example, for parameters $P_1$, a new attractor occurs (Fig. 6) thouroughly different.

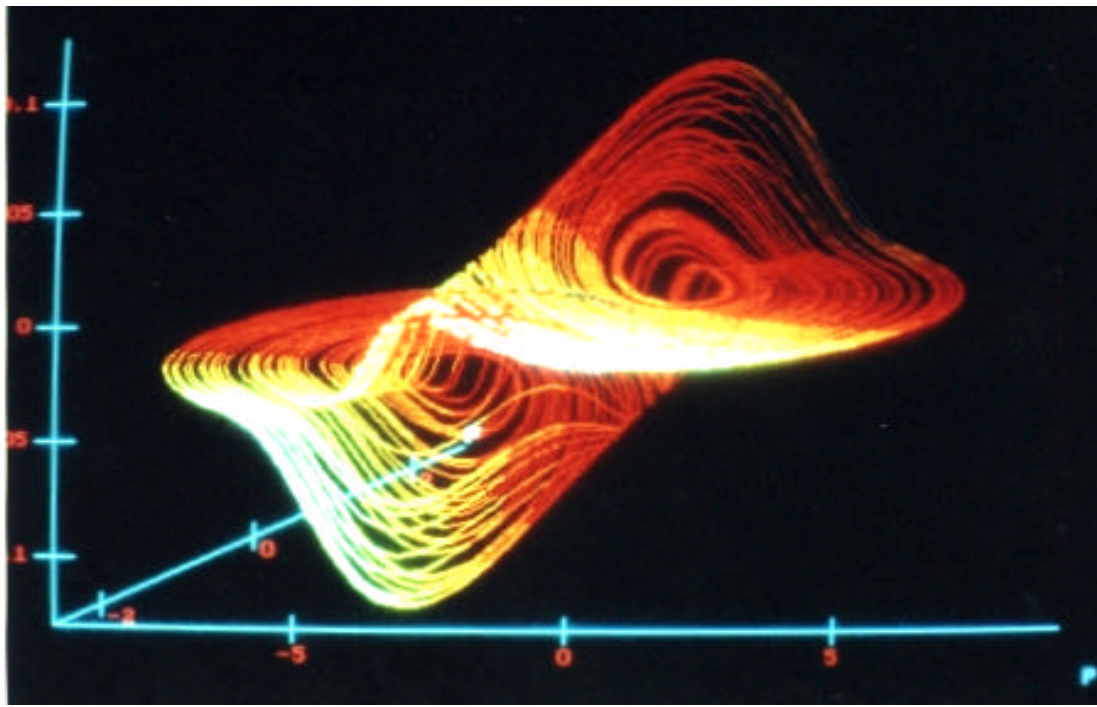

**Fig. 6. Strange attractor for $P_1$ (v, m, n, r, s) = (1.02, 0.02, 0.3, 0.1, 10)**

The total results in 3D -like the Poincaré maps in 2D (Fig. 7)- leads us the analytical investigation of the strange characteristics of this chaos. In fact, the attractors in figures 2 and 4 have different morphologies while a sole parameter has been modified. This propriety represents the Sensitive Dependence on Parameter (SDP).

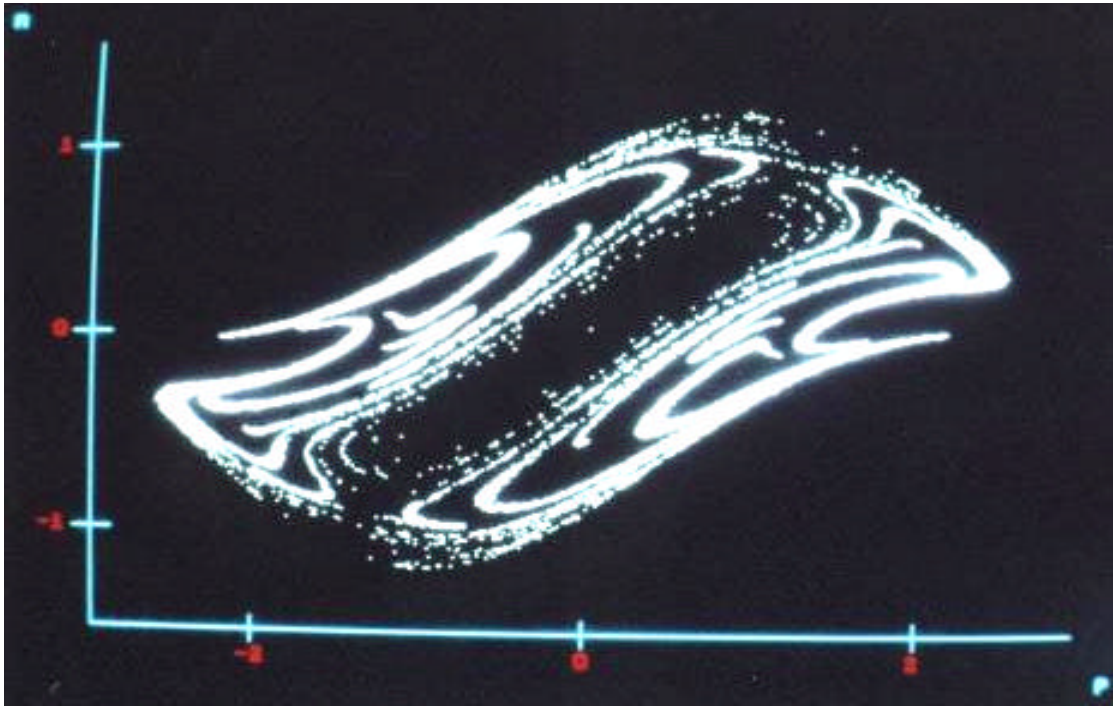

**(a)** The phase portrait of Profits and Reinvestments.

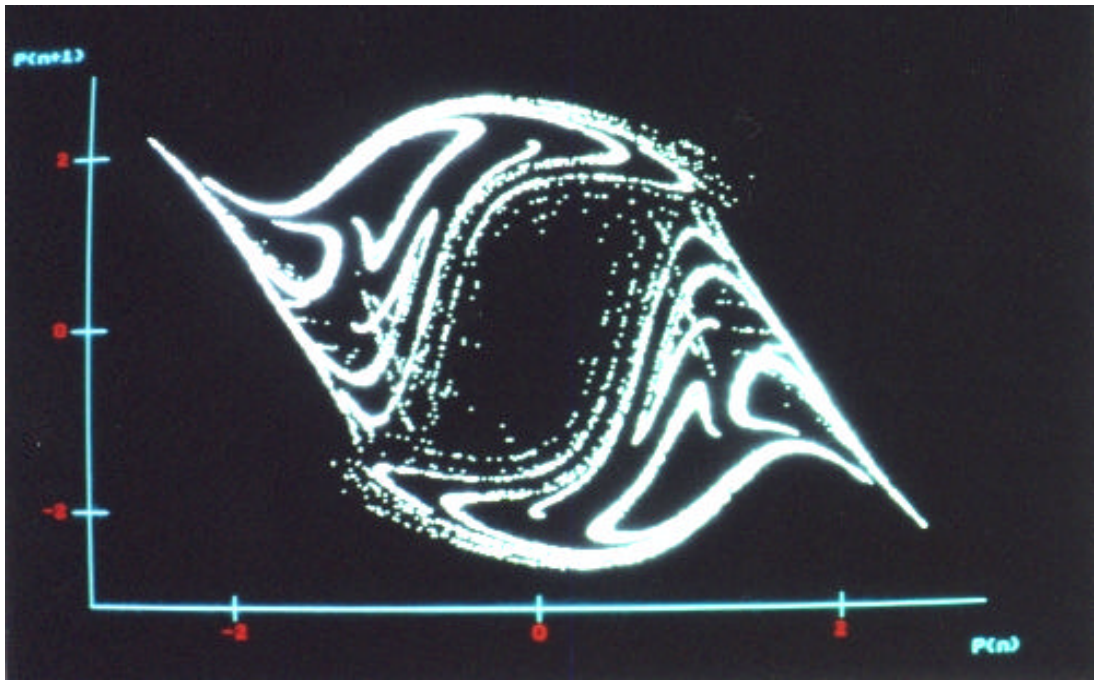

**(b)** The two-dimensional return map of Profits.
**Fig. 7. Poincaré Maps for P$_2$ (v, m, n, r, s) = (7.69, 0.01, 0.05, 0.09, 0.13).**

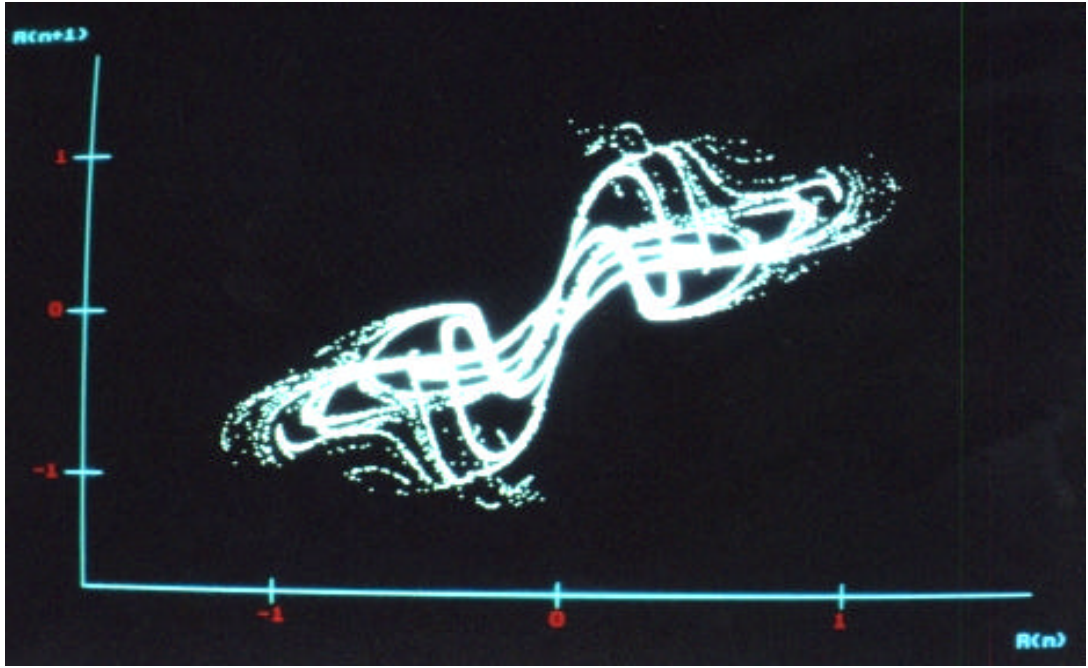

**(c)** The two-dimensional return map of Reinvestments.
**Fig. 7. (continue)**

At another level, restarting the simulation in figure 2 with antisymmetric initial conditions (P, R, F) = (- 0.01, - 0.01, - 0.01) yields a second attractor (Fig. 8) displayed in an independent sub-basin. This result provides an evidence for the Sensitive Dependence on Initial Conditions (SDIC) of the surveyed system.

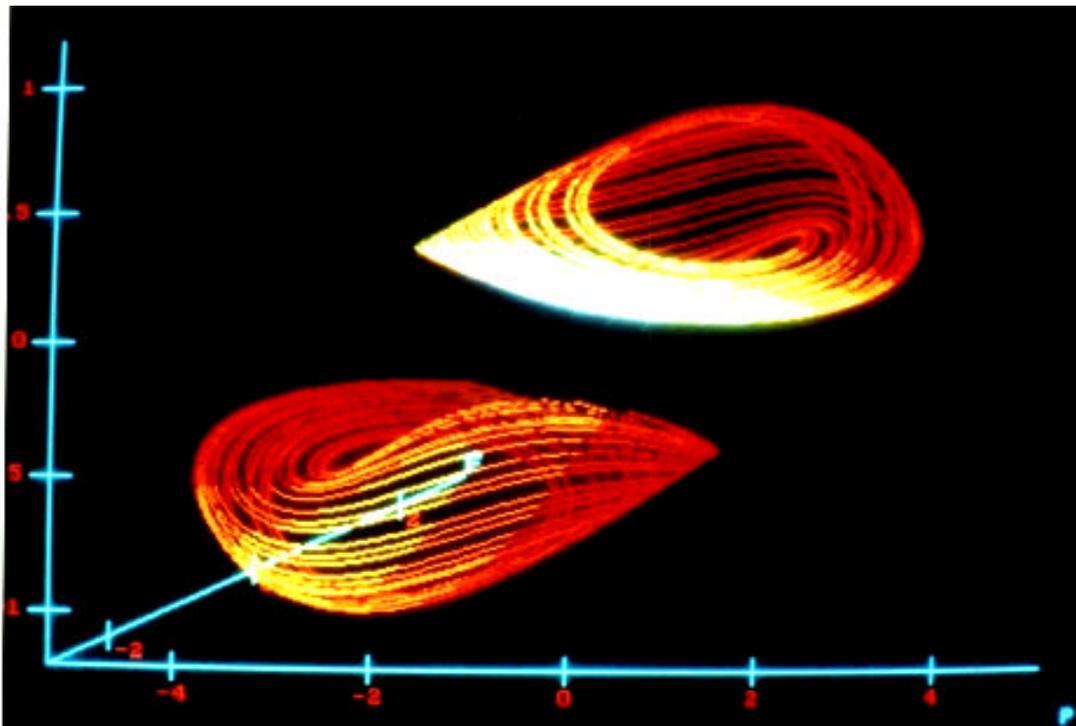

**Fig. 8. Two antisymmetric chaotic attractors in separated sub-basins for $P_0$ and s = 0.3.**

The simultaneous presence of SDP and SDIC testifies the strange features of the attractors. Theoretically, a stationary version of the van der Pol's equation (1926) written in 2 first-order O.D.Es [Thompson and Steward, 1986] constitutes the dynamic foundation of our firm model (Fig. 9).

$$dP/dt = R/v + F/v \quad [1]$$

$$dR/dt = m\,P + n\,(1 - P^2)\,R \quad [2]$$

$$dF/dt = -r\,P + s\,R \quad [3]$$

**Fig. 9. The van der Pol's equation in our system**

In a previous paper [Bouali, 1999], we analyzed the outcome of a retroactive loop (Eq. [3]) determined by the state variables of the van der Pol's oscillator itself. This feedback linkage produces a new class of strange attractors.

### III. Managerial Paradox

Corporate governance have now reached a level of sophistication far beyond our idealized numerical experiments. However, they are a key for a better understanding of the outcome of finance strategies.

Wherein a dynamic perspective, rules and principles of finance governance built in static framework may lose their validity. The findings of a well corporate debt policy connected to a well dividend policy may lead to an umpredictible and hazardous motion of the profit. In our 3D system, the rise of the loss level is an endogenous outcome of the borrowing policy and is not determined by a shock of economic recession. Against the common sense, the profit motion is worsened by the braking of dividend distribution !

The *management art* is the ability to avoid a rigid and irreversible finance program tending to a similar nonlinear turbulence amplification or stall of the growth trend. The best decisions are therefore those which break the hazardous implications for the equilibria of the firm. This art is the choice - within tools of financial management- of a set of policies which allows a flexible guidance of the state variables.

# Appendix

We introduce briefly the fundamental proprieties of the 3D system with only one nonlinear equation:

$$dP/dt = (R + F)/v \quad [1]$$

$$dR/dt = mP + n(1 - P^2)R \quad [2]$$

$$dF/dt = -rP + sR \quad [3]$$

Steady-state equilibria are obtained for $dP/dt = dR/dt = dF/dt = 0$.
We get $R = -F$ from Eq. [1], $R = mP/n(P^2 - 1)$ from Eq. [2] and $P = sR/r$ from Eq. [3]. The last two relations yielded the following equality:
$P[nrP^2 - nr - ms] = 0$. Indeed, the three roots of P are:
$P_1 = 0$, $P_2 = [(nr + ms)/nr]^{1/2}$ and $P_3 = -[(nr + ms)/nr]^{1/2}$.
Let $[(nr + ms)/nr]^{1/2} = k$, the three equilibria become: $E_1 (P, R, F) = (0, 0, 0)$, $E_2(P, R, F) = (k, rk/s, -rk/s)$ and $E_3 (P, R, F) = (-k, -rk/s, rk/s)$. $E_2$ and $E_3$ are antisymmetric.
On the other hand, the Jacobian matrix is:

$$J(P, R, F) = \begin{vmatrix} 0 & 1/v & 1/v \\ m - 2nPR & n(1 - P^2) & 0 \\ -r & s & 0 \end{vmatrix}$$

Hence, we obtain $|J| = [nr(1 - P^2) + ms - 2nPRs]/v$

For all P (v, m, n, r, s) = (4, 0.04, 0.02, 0.1, s), $E_1$ is unstable since $|J| > 0$ for any (positive) value of s parameter.

The two other equilibria, for instance for s = 0.3, are $E_2 = (2.64, 0.88, -0.88)$ and $E_3 = (-2.64, -0.88, 0.88)$. In fact, this s value characterizes two antisymmetric strange attractors. Indeed, two eigenvalues are complex and one is real.

Eventually, the model is conservative for the trajectories that are close to $E_1$ and dissipative for the peripheral orbits ($|P| > 1$), particularly at the neighborhood of $E_2$ and $E_3$; then, the contraction of the volumes by unit time is $\exp.^{0.02(1 - P^2)}$.

# References

**Bouali S.** (1999), Feedback Loop in Extended van der Pol's Equation Applied to an Economic model of Cycles, *Internatioal Journal of Bifurcation and Chaos*, Vol. 9, 4, pp. 745- 756.

**Fama E. F.** and **French K. R.** (1998), Taxes, Financing Decisions, and Firm Value, *Journal of finance*, Vol. 53, pp. 819-844.

**Fluck Z.** (1999), The Dynamics of the Management-Shareholder Conflict, *Review of Financial Studies*, Vol. 11, pp. 347-377.

**Glais M.** (1992), *Economie Industrielle. Les Stratégies Concurrentielles des Firmes.* Edition Litec, Paris.

**Kalay A.** (1980), Signalling, Information Content and the Reluctance to Cut Dividend, *Journal of Financial and Quantitative Analysis,* Vol. XV, 11, pp. 855-784.

**Lintner J.** (1956), Distribution of Incomes of Corporations among Dividends, Retained Earnings and Taxes, *American Economic Review*, Vol. 61, pp. 97-113.

**Medio A.** (1993), *Chaotic Dynamics. Theory and Applications to Economics*, Cambridge University Press, Cambridge.

**Myers S. C.** (1974), Interactions of Corporate Financing and Investment Decisions-Implications for Capital Budgeting, *Journal of Finance*, Vol. XXIX, 1, pp. 1-26.

**Myers S. C.** (1977), Determinants of Corporate Borrowing, *Journal of financial Economics*, Vol. 5, pp. 147-175.

**Pol (van der) B.** (1926), On Relaxation Oscillations, *Phil. Mag.*, Vol. 7, 2, pp. 978- 992.

**Thompson J. M. T. and Stewart H. B.** (1986), *Nonlinear Dynamics and Chaos*, Wiley, New York.